\documentclass[seceqn]{elsart}

\usepackage{amssymb}
\usepackage{amsmath}
\usepackage{bm}

\renewcommand{\Re}{\mathop{\mathrm{Re}}\nolimits}
\renewcommand{\Im}{\mathop{\mathrm{Im}}\nolimits}
\newcommand{\tr}{\mathop{\mathrm{tr}}\nolimits}
\newcommand{\pv}{\mathop{\mathrm{P}}\nolimits}
\newcommand{\ket}[1]{|{#1}\rangle}
\newcommand{\bra}[1]{\langle{#1}|}

\newcommand{\bracket}[2]{\langle#1|#2\rangle}
\newcommand{\rint}{\int\displaylimits}
\newcommand{\tolimit}[2]{\xrightarrow{#1}}
\renewcommand{\d}{d}
\renewcommand{\e}{e}
\newcommand{\openone}{1}

\begin{document}
\begin{frontmatter}



\title{On the assumption of initial factorization in the master equation for weakly coupled systems II: Solvable models}


\author[was2]{K. Yuasa\thanksref{aist}},
\author[was1]{S. Tasaki},
\author[ba1,ba]{P. Facchi},
\author[was2]{G. Kimura\thanksref{tohoku}},
\author[was2]{H. Nakazato},
\author[was2]{I. Ohba},
\author[ba2,ba]{S. Pascazio}

\address[was2]{Department of Physics, Waseda University, Tokyo 169-8555, Japan}
\address[was1]{Department of Applied Physics and Advanced Institute for Complex Systems, Waseda University, Tokyo 169-8555, Japan}
\address[ba1]{Dipartimento di Matematica, Universit\`a di Bari, I-70125 Bari, Italy}
\address[ba]{Istituto Nazionale di Fisica Nucleare, Sezione di Bari, I-70126 Bari, Italy}
\address[ba2]{Dipartimento di Fisica, Universit\`a di Bari, I-70126 Bari, Italy}
\thanks[aist]{Present address: Research Center for Information Security, National Institute of Advanced Industrial Science and Technology (AIST), 1-18-13 Sotokanda, Chiyoda-ku, Tokyo 101-0021, Japan; \textit{E-mail address:} kazuya.yuasa@aist.go.jp}
\thanks[tohoku]{Present address: Graduate School of Information Sciences, Tohoku University, Sendai 980-8579, Japan.}

\date{27 July 2006}

\begin{abstract}
We analyze some solvable models of a quantum mechanical system in
interaction with a reservoir when the initial state is not
factorized. We apply Nakajima--Zwanzig's projection method by
choosing a reference state of the reservoir endowed with the mixing
property. In van Hove's limit, the dynamics is described in terms of
a master equation. We observe that Markovianity becomes a valid
approximation for timescales that depend both on the form factors of
the interaction and on the observables of the reservoir that can be
measured.
\end{abstract}


\end{frontmatter}


\section{Introduction}
The dissipative dynamics of a small quantum system weakly coupled to
a large reservoir is described in terms of a master equation
\cite{ref:SpohnReview,ref:KuboTextbook,ref:WeissTextbook,ref:QuantumNoise}.
In the standard approach to this problem, one usually takes for
granted that there are no initial correlations between the system
and the reservoir. In the preceding article \cite{ref:ArticleI}, hereafter referred to
as Article I, we  reconsidered this hypothesis in the framework of
 Nakajima--Zwanzig's projection method
\cite{ref:KuboTextbook,ref:QuantumNoise,ref:Projection,ref:HaakeGeneralizedMasterEq}
and proved that, in order to get a consistent description, the
reference state of the reservoir should be endowed with the mixing
property. In such a case, the initial correlations disappear in
the Markovian (van Hove) limit and the system behaves as if it
started from a factorized initial condition. Interestingly, one
arrives at the same conclusions also for uncorrelated initial
conditions. The mixing property is therefore crucial, and a
``wrong'' choice of the reservoir state provokes the appearance of
secular terms.

In this article, we shall focus on the hypotheses that are necessary
for the derivation of the theorem proved in Article I \cite{ref:ArticleI}. These will
be scrutinized in terms of two exactly solvable models, in which an
oscillator is coupled to a bosonic reservoir. This will enable us to
describe the onset to Markovianity and the timescales at which
Markovianity becomes a valid approximation.

This article is organized as follows. We introduce notation and
summarize previous results in Sec.\ \ref{sec:general}. The first
exactly solvable model is introduced in Sec.\
\ref{sec:ExactlySolvableModel} and solved in Secs.\
\ref{sec:correla}--\ref{sec:solution}.
The second model is briefly discussed in Sec.\
\ref{sec:model2}.
Section \ref{sec:Summary} is devoted to a discussion and some
concluding remarks. Two Appendices contain the details of the
derivations.

\section{Summary of Previous Results}
\label{sec:general}
\subsection{Notation}
We start by briefly summarizing the main ideas of Article I \cite{ref:ArticleI} and
introduce notation. Let the total system consist of a ``large''
reservoir B and a ``small'' (sub)system S, so that the total
Hilbert space can be expressed as the tensor product of the
Hilbert spaces of the reservoir $\mathcal{H}_\mathrm{B}$ and of
the system $\mathcal{H}_\mathrm{S}$,
\begin{equation}
\mathcal{H}_\mathrm{tot}
=\mathcal{H}_\mathrm{S}\otimes\mathcal{H}_\mathrm{B} .
\label{tothilb}
\end{equation}
The Hamiltonian and the corresponding Liouvillian of the total
system read
\begin{equation}
H = H_0 + \lambda H_\mathrm{SB} = H_\mathrm{S}+ H_\mathrm{B}+\lambda
H_\mathrm{SB} ,\label{Hamiltonian1}
\end{equation}
\begin{equation}
\mathcal{L}= \mathcal{L}_0 + \lambda \mathcal{L}_\mathrm{SB}
=\mathcal{L}_\mathrm{S}+\mathcal{L}_\mathrm{B} + \lambda
\mathcal{L}_\mathrm{SB} ,
\label{Liou1}
\end{equation}
respectively, where $\lambda$ is the coupling constant. Clearly,
\begin{equation}
[H_\mathrm{S}, H_\mathrm{B}] =0 , \quad
[\mathcal{L}_\mathrm{S},\mathcal{L}_\mathrm{B}]=0.
\label{Lioucomm}
\end{equation}
We assume that the system Hamiltonian $H_\mathrm{S}$ admits a pure
point spectrum, and the system Liouvillian $\mathcal{L}_\mathrm{S}$
is resolved in terms of its eigenprojections $\tilde{Q}_m$,
\begin{equation}
\mathcal{L}_\mathrm{S}=-i \sum_m \omega_m\tilde{Q}_m,\quad
\sum_m\tilde{Q}_m=1,\quad
\tilde{Q}_m\tilde{Q}_{n}=\delta_{mn}\tilde{Q}_m.
\label{eqn:eigenA9bis}
\end{equation}

\subsection{Nakajima--Zwanzig's Projection Method}
\label{sec:dynamics}
Let $\rho(t)$ be the density operator of the total system at time
$t$, which has evolved from the initial state $\rho_0$
\begin{equation}
\rho(t)=\e^{\mathcal{L}t}\rho_0
\end{equation}
and is the solution of the von Neumann equation
\begin{equation}
\frac{\d}{\d t}\rho(t)=\mathcal{L}\rho(t),\qquad
\rho(0)=\rho_0.
\label{eqn:A1}
\end{equation}
We are interested in the reduced dynamics of system S, which is
described by the density operator of S,
\begin{equation}\label{eqn:sigtrace}
\rho_\mathrm{S}(t)=\tr_\mathrm{B}\rho(t).
\end{equation}
In order to derive a master equation for $\rho_\mathrm{S}(t)$,
Nakajima--Zwanzig's procedure makes use of the projection operators
\cite{ref:KuboTextbook,ref:QuantumNoise,ref:Projection,ref:HaakeGeneralizedMasterEq}
\begin{equation}
\label{eqn:gendefproj}
\mathcal{P}\rho=\tr_\mathrm{B}\{\rho\}\otimes\Omega_\mathrm{B}
=\sigma\otimes\Omega_\mathrm{B},\qquad
\mathcal{Q}=1-\mathcal{P},
\end{equation}
where
$\Omega_\mathrm{B}$ is a certain \textit{reference state} of the
reservoir. Due to normalization $\tr_\mathrm{B}\Omega_\mathrm{B}=1$,
it follows that $\mathcal{P}^2=\mathcal{P}$ and
$\mathcal{Q}^2=\mathcal{Q}$. In particular,
\begin{equation}
\label{eqn:defproj} \mathcal{P}\rho(t)
=\rho_\mathrm{S}(t)\otimes\Omega_\mathrm{B},\qquad
\mathcal{Q}\rho(t)=\rho(t)-\rho_\mathrm{S}(t)\otimes\Omega_\mathrm{B},
\end{equation}
where we used the definition (\ref{eqn:sigtrace}).

In the standard derivation of a master equation, the initial state
of the total system, $\rho_0$, is taken to be the tensor product of
a system initial state $\rho_\mathrm{S}$ and a reservoir state
$\rho_\mathrm{B}$,
\begin{equation}\label{eqn:rhoprod}
\rho_0=\rho_\mathrm{S}\otimes\rho_\mathrm{B}.
\end{equation}
This is an \textit{uncorrelated} initial state. The reservoir is
assumed to be at equilibrium (with respect to the reservoir free
evolution $\mathcal{L}_\mathrm{B}$)
\begin{equation}\label{BathEq}
\mathcal{L}_\mathrm{B}\rho_\mathrm{B}=0,
\end{equation}
and in most applications $\rho_\mathrm{B}=Z_\beta^{-1}\e^{-\beta
H_\mathrm{B}}$ is a thermal state at the inverse temperature
$\beta=(k_\mathrm{B}T)^{-1}$ with the normalization constant
$Z_\beta$. Then, the reservoir state $\rho_\mathrm{B}$ in the
uncorrelated initial state (\ref{eqn:rhoprod}) is usually taken as
the reference state $\Omega_\mathrm{B}$.

When the assumption of a factorized initial state is not justified,
however, an ambiguity arises regarding the choice of the reference
state $\Omega_\mathrm{B}$. Indeed, if
\begin{equation}
\rho_0=\rho_\mathrm{S}\otimes\rho_\mathrm{B}+\delta\rho_0,
\label{eqn:CanonicalFormState}
\end{equation}
where
\begin{equation}
\rho_\mathrm{S}=\tr_\mathrm{B}\rho_0,\qquad
\rho_\mathrm{B}=\tr_\mathrm{S}\rho_0,
\end{equation}
and the term $\delta\rho_0$ represents the correlation  between
system S and reservoir B, the relation between $\rho_\mathrm{B}$ and
$\Omega_\mathrm{B}$ is by no means obvious. We discussed this point
in Article I \cite{ref:ArticleI} and proved the following theorem.

\subsection{Theorem}
\label{sec:theoremmain}
Given a correlated initial state $\rho_0$, if
\begin{enumerate}
\renewcommand{\labelenumi}{(\roman{enumi})}
\item \label{spectrum} $0$ is the unique simple eigenvalue of the
reservoir Liouvillian $\mathcal{L}_\mathrm{B}$ corresponding to the
eigenvector $\Omega_\mathrm{B}$ and the remaining part of the
spectrum of $\mathcal{L}_\mathrm{B}$ is absolutely continuous
(strictly speaking, the spectrum of $\mathcal{L}_\mathrm{B}$ can be
defined only once the sector has been specified: in our case, the
relevant sector is that containing the state $\Omega_\mathrm{B}$);
\item the initial (correlated) state of the total system is given
 in the form
\begin{equation}
\rho_0=\Lambda(\openone_\mathrm{S}\otimes\Omega_\mathrm{B})
=\sum_iL_i(\openone_\mathrm{S}\otimes\Omega_\mathrm{B})L_i^\dag,
\label{eqn:CondInitialState}
\end{equation}
where $\Lambda$ is a bounded superoperator (i.e., $L_i$'s are
bounded operators) satisfying the normalization condition
$\tr\rho_0=1$, namely, the initial state $\rho_0$ belongs to the
sector specified by $\openone_\mathrm{S}\otimes\Omega_\mathrm{B}$,
\end{enumerate}
then van Hove's ``$\lambda^2 t$'' limit
\cite{ref:SpohnReview,ref:VanHoveLimit,ref:FacchiPascazioVanHovePhysica}
of the $\mathcal{P}$-projected density operator in the interaction
picture,
\begin{equation}\label{eqn:densdef}
\rho_{\mathrm{I}}(\tau) = \lim_{\lambda \to 0}
\rho_\mathrm{I}^{(\lambda)}(\tau) =
 \lim_{\lambda \to 0} \e^{-\mathcal{L}_\mathrm{S}\tau/\lambda^2}
\mathcal{P} \rho(\tau/\lambda^2),
\end{equation}
is the solution of
\begin{equation}\label{eqn:inteq}
\rho_\mathrm{I}(\tau)=\mathcal{P}\rho_0
+\rint_0^\tau \d\tau'\,\mathcal{K}\rho_\mathrm{I}(\tau')
\end{equation}
with
\begin{equation}
\mathcal{K}
=-\sum_m\mathcal{P}\tilde{Q}_m\mathcal{L}_\mathrm{SB}
\frac{\mathcal{Q}}{\mathcal{L}_0+i\omega_m-0^+}
\mathcal{L}_\mathrm{SB}\tilde{Q}_m\mathcal{P},
\label{eqn:Kdef}
\end{equation}
or equivalently,
\begin{equation}\label{eqn:diffeq}
\frac{\d}{\d\tau}\rho_\mathrm{I}(\tau)
=\mathcal{K}\rho_\mathrm{I}(\tau),\qquad
\rho_\mathrm{I}(0) =\mathcal{P}\rho_0
=\tr_\mathrm{B}\{\rho_0\}\otimes\Omega_\mathrm{B}.
\end{equation}
That is, even if the initial state $\rho_0$ is not in a factorized
form, all correlations disappear in van Hove's limit and system S
behaves as if the total system started from the factorized initial
state in (\ref{eqn:diffeq}) with the reservoir state
$\Omega_\mathrm{B}$.

In addition, we showed that
\begin{equation}
\lim_{\lambda\to0}\mathcal{Q}\rho(\tau/\lambda^2)=0,
\label{eqn:Theorem2}
\end{equation}
which makes the dynamics consistent, for no spurious term will
develop in the master equation and no correlations can appear at
later times: not only the initial state, but also the state at any
moment $t$ is factorized in van Hove's limit. This supports the
validity of the assumption of the factorized state, that is
frequently applied in literature in order to derive a master
equation
\cite{ref:KuboTextbook,ref:WeissTextbook,ref:QuantumNoise}. The
state of system S evolves according to the master equation
(\ref{eqn:diffeq}), while the reservoir B remains in the state
$\Omega_\mathrm{B}$.

It is important to note that, in van Hove's limit, the reservoir
state immediately relaxes into $\Omega_\mathrm{B}$, which is the
eigenstate of the reservoir Liouvillian $\mathcal{L}_\mathrm{B}$
belonging to its unique simple eigenvalue $0$, and the spectral
properties required in hypothesis (i) imply that it is a
\textit{mixing} state. The right choice for the reference state of
the projection (\ref{eqn:defproj}) is this mixing state
$\Omega_\mathrm{B}$, and such a projection is nothing but the
eigenprojection of the reservoir Liouvillian
$\mathcal{L}_\mathrm{B}$ belonging to the simple eigenvalue $0$.
This is the criterion for the reference state, that covers
both equilibrium states and nonequilibrium steady states.
Furthermore, as clarified in Article I \cite{ref:ArticleI}, the interaction between
system S and reservoir B is not essential to the factorization or
the mixing; the total system is factorized and the reservoir relaxes
into the mixing state through its own free evolution.

The purpose of the present article is to scrutinize these issues
in some explicit examples. In particular, we shall focus on: (a)
the disappearance of the initial correlation, (b) the
factorization of the total system, and (c) the relaxation of the
reservoir into the mixing state, in van Hove's limit. This will
also enable us to discuss the relevant timescales for the
factorization and the mixing.

\section{An Exactly Solvable Model}\label{sec:ExactlySolvableModel}
Let us corroborate the above general arguments by scrutinizing an
exactly solvable model. We consider an oscillator $a$ coupled to a
reservoir $b_\omega$, whose Hamiltonian is given by
(\ref{Hamiltonian1}) with
\begin{equation}
H_\mathrm{S}=\omega_\mathrm{S} a^\dag a,\ \ %
H_\mathrm{B}=\rint_0^\infty \d\omega\,\omega
b_\omega^\dag b_\omega,\ \ %
H_\mathrm{SB}=i\rint_0^\infty \d\omega\,
(g_\omega^*a^\dag b_\omega-g_\omega ab_\omega^\dag),
\label{eqn:SolvableHamiltonian}
\end{equation}
where $a$ ($a^\dag$) and $b_\omega$ ($b_\omega^\dag$) are
annihilation (creation) operators satisfying the canonical
commutation relations
\begin{equation}
[a,a^\dag]=1,\qquad
[b_\omega,b_{\omega'}^\dag]=\delta(\omega-\omega'),
\label{eqn:CCR}
\end{equation}
and $g_\omega$ is the form factor of the interaction. Even though
system S has an infinite number of levels, and does not fulfill
the conditions of the main theorem proved in Article I \cite{ref:ArticleI}, the
following explicit calculation will show that all the conclusions
are still valid and therefore the theorem has a wider
applicability.

The above model is exactly solvable
\cite{ref:FordLewisOConnellQuantumLangevin,ref:FordOConnellAutoCorrelation2000,ref:Caldeira2000}.
Indeed, the Heisenberg equations of motion for $a(t)=\e^{i
Ht}a\e^{-i Ht}$ and $b_\omega(t)=\e^{i Ht}b_\omega\e^{-i Ht}$
read
\begin{subequations}
\begin{gather}
\dot{a}(t)=-i\omega_\mathrm{S} a(t)
+\lambda\rint_0^\infty\d\omega\,g_\omega^*b_\omega(t),\\
\dot{b}_\omega(t)=-i\omega b_\omega(t)
-\lambda g_\omega a(t),
\end{gather}
\end{subequations}
and by integrating the second equation and inserting it into the
first, one obtains an integro-differential equation for $a(t)$,
\begin{equation}
\dot{a}(t)=-i\omega_\mathrm{S} a(t)-\lambda^2\rint_0^t\d t'\,K(t-t')a(t')
+\lambda B(t) ,
\end{equation}
with
\begin{equation}
K(t)=\rint_0^\infty\d\omega\,|g_\omega|^2\e^{-i\omega t},\qquad
B(t)=\rint_0^\infty\d\omega\,g_\omega^*\e^{-i\omega t}b_\omega,
\label{eqn:MemoryNoise}
\end{equation}
which is solved via Laplace transform to yield
\begin{subequations}
\label{eqn:HeisenbergSolution}
\begin{gather}
a(t)=G(t)a+\lambda\rint_0^t\d t'\,G(t-t')B(t'),\\
b_\omega(t)=\e^{-i\omega t}b_\omega
-\lambda\rint_0^t\d t'\,\e^{-i\omega(t-t')}g_\omega a(t'),
\end{gather}
\end{subequations}
where
\begin{equation}
G(t)=\rint_{C_\text{B}}\frac{\d s}{2\pi i}
\frac{\e^{st}}{s+i\omega_\mathrm{S}+\lambda^2\hat{K}(s)},\qquad
\hat{K}(s)=\rint_0^\infty\d\omega\,\frac{|g_\omega|^2}{s+i\omega},
\label{eqn:G}
\end{equation}
$C_\text{B}$ being the Bromwich path on the complex $s$-plane.
Note that $G(0^+)=1$ and $\dot{G}(0^+)=-i\omega_\mathrm{S}$.

\section{A Correlated Initial State}\label{sec:correla}
Any physical preparation of a quantum state is based on concrete
physical procedures that cannot be controlled with complete
accuracy. The real initial state is therefore unknown to some
extent and in general has certainly some correlations built in. As
an example of a correlated initial state, that has the advantage
of being solvable, we take
\begin{equation}
\rho_0=\frac{1}{Z_0}\e^{a^\dag\xi^\dag b}(\sigma_\text{S}\otimes\rho_\mathcal{W})\e^{b^\dag\xi
a},
\label{eqn:InitialState}
\end{equation}
with any positive operator $\sigma_\text{S}$ of system S and a
reservoir state
\begin{equation}
\rho_\mathcal{W}=\frac{1}{Z_\mathcal{W}}\e^{-b^\dag\mathcal{W}b},
\label{eqn:GaussianState}
\end{equation}
where the summations over the reservoir modes $\omega$ are implicit
(and so henceforth as long as no confusion can arise):
\begin{equation}
\xi^\dag b=\rint_0^\infty\d\omega\,\xi_\omega^*b_\omega,\qquad
b^\dag\mathcal{W}b=\rint_0^\infty\d\omega\,\rint_0^\infty\d\omega'\,
b_\omega^\dag\mathcal{W}_{\omega\omega'}b_{\omega'}.
\end{equation}
$\mathcal{W}_{\omega\omega'}$ is Hermitian
($\mathcal{W}_{\omega\omega'}= \mathcal{W}_{\omega'\omega}^*$) and
consists of $\mathcal{W}_{\omega\omega'}^{(0)}$, that is
proportional to $\delta(\omega-\omega')$, and the remaining square
integrable part $\tilde{\mathcal{W}}_{\omega\omega'}$,
\begin{equation}
\mathcal{W}_{\omega\omega'} =\mathcal{W}_{\omega\omega'}^{(0)}
+\tilde{\mathcal{W}}_{\omega\omega'},\qquad
\mathcal{W}_{\omega\omega'}^{(0)}=W(\omega)\delta(\omega-\omega').
\label{eqn:Wdecomp}
\end{equation}
The states $\rho_0$ and $\rho_\mathcal{W}$ are normalized with the
normalization constants $Z_0$ and $Z_\mathcal{W}$, and $\xi_\omega$
is the relevant parameter to the initial correlation between system
S and reservoir B\@.

For $\xi_\omega=0$, the state (\ref{eqn:InitialState}) is obviously
factorized, while it becomes a tightly correlated state for any
$\xi_\omega\neq0$, with correlations proportional to $\xi_\omega$, as will be
shown later in (\ref{eqn:SBcorrelation}). Actually, the operator
$\e^{a^\dag\xi^\dag b}$ appearing in the initial state
(\ref{eqn:InitialState}) generates a correlation between S and B: it
changes the $n$-particle states $b^\dag\eta_1\cdots
b^\dag\eta_n|\mathrm{vac}\rangle$ of the reservoir into $\e^{a^\dag\xi^\dag b}b^\dag\eta_1\cdots
b^\dag\eta_n|\mathrm{vac}\rangle
=(b^\dag\eta_1+a^\dag\xi^\dag\eta_1)\cdots(b^\dag\eta_n+a^\dag\xi^\dag\eta_n)|\mathrm{vac}\rangle$,
so that S and B are entangled for any nonvanishing value of
$\xi_\omega$. It is also possible to explicitly compute the
correlation functions in the initial state (\ref{eqn:InitialState}):
see the generating functional (\ref{eqn:CharFuncInit}) and the
correlation function (\ref{eqn:SBcorrelation}) below. The choice of
this particular form for the initial state $\rho_0$ is mainly due to
the fact that it allows us to solve the dynamics of the total system
exactly and to discuss the correlation between system S and
reservoir B\@. One can think of the correlations in
(\ref{eqn:InitialState}) as engendered by a linear interaction of
the form $H_{\mathrm{prep}}\propto a^\dag
\xi^\dag b + \mathrm{h.c.}$ in a rotating-wave-like
approximation.

As shown in Appendix A of Article I \cite{ref:ArticleI}, the (normalized) reservoir
state
\begin{equation}
\Omega_\mathrm{B}
=\frac{1}{Z_{\mathcal{W}_0}}\e^{-b^\dag\mathcal{W}^{(0)}b}
\label{eqn:MixingModel}
\end{equation}
is mixing with respect to the reservoir dynamics driven by the
Hamiltonian $H_\mathrm{B}$ in (\ref{eqn:SolvableHamiltonian}), and
the initial state $\rho_0$ in (\ref{eqn:InitialState}) belongs to
the sector specified by
$\openone_\mathrm{S}\otimes\Omega_\mathrm{B}$ in the sense of
(\ref{eqn:CondInitialState}). Indeed, $\rho_0$ is the state
perturbed from $\openone_\mathrm{S}\otimes\Omega_\mathrm{B}$ by a
local operator $L$,
\begin{equation}
\rho_0
=L(\openone_\mathrm{S}\otimes\Omega_\mathrm{B})L^\dag,\qquad
L=\frac{1}{\sqrt{Z_0}}\e^{a^\dag\xi^\dag b}
(\sqrt{\sigma_\text{S}}\otimes L_\text{B}),
\label{eqn:EquiSectorModel}
\end{equation}
where
\begin{equation}
L_\mathrm{B}
=\rho_\mathcal{W}^{1/2}\Omega_\mathrm{B}^{-1/2}
=\sqrt{\frac{Z_{\mathcal{W}_0}}{Z_\mathcal{W}}}
\mathop{\bar{\mathrm{T}}}\exp\Biggl( -\rint_0^{1/2}\d\beta\,
b^\dag \e^{-\beta\mathcal{W}^{(0)}}\tilde{\mathcal{W}}
\e^{\beta\mathcal{W}^{(0)}}b \Biggr)
\label{eqn:LB}
\end{equation}
is a local perturbation such that
\begin{equation}
\rho_\mathcal{W}
=L_\mathrm{B}\Omega_\mathrm{B}L_\mathrm{B}^\dag ,
\label{eqn:RhoW_OmegaB}
\end{equation}
$\mathop{\bar{\mathrm{T}}}$ denoting the anti-chronologically
ordered product and
\begin{equation}
b^\dag\e^{-\beta\mathcal{W}^{(0)}}\tilde{\mathcal{W}}
\e^{\beta\mathcal{W}^{(0)}}b =\rint_0^\infty\d\omega\rint_0^\infty\d\omega'\,
b_\omega^\dag\e^{-\beta
W(\omega)}\tilde{\mathcal{W}}_{\omega\omega'} \e^{\beta
W(\omega')}b_{\omega'}.
\end{equation}
Even though the initial state $\rho_0$ does not satisfy the
hypotheses of the theorem proved in Article I \cite{ref:ArticleI}, the following
analysis extends the general results valid for a bounded
perturbation.

Note that the reservoir Gaussian state $\rho_\mathcal{W}$ in
(\ref{eqn:GaussianState}) is fully characterized by the two-point
function
\begin{equation}
\mathcal{N}_{\omega\omega'}
=\langle b_{\omega'}^\dag b_\omega\rangle_{\rho_\mathcal{W}}
=\tr_\text{B}\{b_{\omega'}^\dag b_\omega\rho_\mathcal{W}\}
\label{eqn:ExtBoseFunc}
\end{equation}
and, as shown in Appendix A in Article I \cite{ref:ArticleI}, it is also composed of two
parts like $\mathcal{W}_{\omega\omega'}$ in (\ref{eqn:Wdecomp}),
\begin{equation}
\mathcal{N}_{\omega\omega'}
=\mathcal{N}^{(0)}_{\omega\omega'}
+\tilde{\mathcal{N}}_{\omega\omega'}.
\label{eqn:Ndecomp}
\end{equation}
The first term is the two-point function in the mixing state
$\Omega_\mathrm{B}$,
\begin{equation}
\mathcal{N}_{\omega\omega'}^{(0)}
=\langle b_{\omega'}^\dag b_\omega\rangle_{\Omega_\mathrm{B}}
=N(\omega)\delta(\omega-\omega'),\quad
N(\omega)=\frac{1}{\e^{W(\omega)}-1},
\label{eqn:BoseFunc}
\end{equation}
which is the Bose distribution function when
$W(\omega)=\beta\omega$, while the second one is a local function
representing the effect of the local perturbation $L_\mathrm{B}$
in (\ref{eqn:RhoW_OmegaB}).

\section{Dynamics of the Total System}
Since we are interested in the correlation between system S and
reservoir B, we need to look at the state of the \textit{total}
system, $\rho(t)$. In order to treat the reservoir degrees of
freedom rigorously, we should restrict ourselves to reservoir
observables whose expectation values are finite and discuss the
state of the total system through a set of such expectation values.
The relevant quantity for our discussion is therefore a
characteristic functional of the state $\rho(t)$, e.g.\
\begin{equation}
\mathcal{G}[J_a,J_a^*,J_b,J_b^\dag;t]
=\tr\{
\e^{J_a^*a}\e^{J_b^\dag b}\e^{-b^\dag J_b}\e^{-a^\dag J_a}
\rho(t)\},
\label{eqn:CharFuncDef}
\end{equation}
where $J_b^\dag b=\rint_0^\infty\d\omega\,J_{b,\omega}^*b_\omega$,
which is the generating functional of the expectation values of
any anti-normally ordered products of $a$, $a^\dag$, $b_\omega$,
and $b_\omega^\dag$ and characterizes the state of the total
system, $\rho(t)$. It is important to note that we are not
interested in infinitely extended objects, such as the Hamiltonian
of the reservoir $H_\mathrm{B}$, since their expectation values
are infinite: our targets are \textit{locally} distributed
objects. Such a formalization of the problem is reasonable, since
we cannot observe infinitely extended objects in practice, and
this is nothing but the starting point of the $C^*$-algebraic
approach to quantum statistical mechanics \cite{BR}. In the
characteristic functional (\ref{eqn:CharFuncDef}), the bandwidth
of $J_{b,\omega}$ represents the locality of the observables.

Let us begin with the characteristic functional of the initial state
$\rho_0$ in (\ref{eqn:InitialState}),
\begin{equation}
\mathcal{G}_0[J_a,J_a^*,J_b,J_b^\dag]
=\mathcal{G}[J_a,J_a^*,J_b,J_b^\dag;0]
=\tr\{
\e^{J_a^*a}\e^{J_b^\dag b}\e^{-b^\dag J_b}\e^{-a^\dag J_a}
\rho_0\},
\end{equation}
which, in the coherent-state representation ($Q$-representation
\cite{ref:QuantumNoise})
\begin{equation}
a\ket{\alpha}=\alpha\ket{\alpha},\quad
\bracket{\alpha}{\alpha'}
=\e^{-|\alpha|^2/2-|\alpha'|^2/2+\alpha^*\alpha'},\quad
\int\frac{\d^2\alpha}{\pi}\,\ket{\alpha}\bra{\alpha}
=\openone_\text{S},
\end{equation}
is evaluated as
\begin{eqnarray}
&&\mathcal{G}_0[J_a,J_a^*,J_b,J_b^\dag]\nonumber\\
&&\qquad=\frac{1}{Z_0}\e^{-J_b^\dag J_b}
\int\frac{\d^2\alpha}{\pi}\,\bra{\alpha}\sigma_\text{S}\ket{\alpha}
\e^{J_a^*\alpha-\alpha^*J_a}
\langle
\e^{b^\dag(\xi\alpha-J_b)}\e^{(\alpha^*\xi^\dag+J_b^\dag)b}
\rangle_\mathcal{W}\nonumber\\
&&\qquad=\frac{1}{Z_0}\e^{-J_b^\dag(1+\mathcal{N})J_b}
\int\frac{\d^2\alpha}{\pi}\,\bra{\alpha}\sigma_\text{S}\ket{\alpha}
\e^{\alpha^*\xi^\dag\mathcal{N}\xi\alpha}
\e^{(J_a^*+J_b^\dag\mathcal{N}\xi)\alpha}
\e^{-\alpha^*(J_a+\xi^\dag\mathcal{N}J_b)}\nonumber\\
&&\qquad=\e^{-J_b^\dag(1+\mathcal{N})J_b}
\mathcal{G}_\text{S}
(J_a+\xi^\dag\mathcal{N}J_b,
J_a^*+J_b^\dag\mathcal{N}\xi),
\label{eqn:CharFuncInit}
\end{eqnarray}
where
\begin{equation}
\mathcal{G}_\text{S}(J_a,J_a^*)
=\mathcal{G}_0[J_a,J_a^*,0,0]
=\tr_\text{S}\{\e^{J_a^*a}\e^{-a^\dag J_a}\rho_\text{S}\}
\end{equation}
is the characteristic function of the initial state of system S
and
\begin{equation}
\rho_\text{S}=\tr_\text{B}\rho_0.
\end{equation}
One can see from this characteristic functional how the parameter
$\xi_\omega$ embodies the initial correlation. For example,
\begin{eqnarray}
\langle ab_\omega^\dag\rangle_{\rho_0}
&=&\left.
-\frac{\partial}{\partial J_a^*}
\frac{\delta}{\delta J_{b,\omega}}
\mathcal{G}_0[J_a,J_a^*,J_b,J_b^\dag]
\right|_{J_a,J_a^*,J_b,J_b^\dag=0}\nonumber\\
&=&-\rint_0^\infty\d\omega'\,
\xi_{\omega'}^*\mathcal{N}_{\omega'\omega}
\left.
\frac{\partial^2\mathcal{G}_\text{S}(J_a,J_a^*)}{\partial J_a\,\partial J_a^*}
\right|_{J_a,J_a^*=0}
=\rint_0^\infty\d\omega'\,
\xi_{\omega'}^*\mathcal{N}_{\omega'\omega}
\langle aa^\dag\rangle_{\rho_\text{S}}.\nonumber\\
\label{eqn:SBcorrelation}
\end{eqnarray}

Let us discuss the evolution of the state of the total system
\begin{equation}
\rho(t)=\e^{-i Ht}\rho_0\e^{i Ht}.
\end{equation}
The characteristic functional (\ref{eqn:CharFuncDef}) of the state
$\rho(t)$ is easily computed in the Heisenberg picture
\begin{eqnarray}
\mathcal{G}[J_a,J_a^*,J_b,J_b^\dag;t]
&=&\tr\{
\e^{J_a^*a(t)}\e^{J_b^\dag b(t)}\e^{-b^\dag(t)J_b}\e^{-a^\dag(t)J_a}\rho_0
\}\nonumber\\
&=&\tr\{
\e^{J_a^*(t)a}\e^{J_b^\dag(t)b}\e^{-b^\dag J_b(t)}\e^{-a^\dag J_a(t)}\rho_0
\}\nonumber\\
&=&\mathcal{G}_0[J_a(t),J_a^*(t),J_b(t),J_b^\dag(t)],
\label{eqn:CharFunc0}
\end{eqnarray}
where $J_a(t)$ and $J_b(t)$ are functionals of $J_a$ and $J_b$,
defined via $a^\dag(t)J_a+b^\dag(t)J_b=a^\dag J_a(t)+b^\dag J_b(t)$.
Note that the solutions (\ref{eqn:HeisenbergSolution}) for $a(t)$
and $b_\omega(t)$ are linear in $a$ and $b_\omega$, but do not
contain $a^\dag$ or $b_\omega^\dag$. The characteristic functional
of the initial state $\rho_0$ is given in (\ref{eqn:CharFuncInit})
and Eq.~(\ref{eqn:CharFunc0}) is further reduced to
\begin{equation}
\mathcal{G}[J_a,J_a^*,J_b,J_b^\dag;t]
=\e^{-J_b^\dag(t)(1+\mathcal{N})J_b(t)}
\mathcal{G}_\text{S}\bm{(}
J_a(t)+\xi^\dag\mathcal{N}J_b(t),
J_a^*(t)+J_b^\dag(t)\mathcal{N}\xi
\bm{)}.
\label{eqn:CharFunc}
\end{equation}
We thus obtain the exact characteristic functional of the state of
the \textit{total} system, $\rho(t)$,
\begin{equation}
\mathcal{G}[J_a,J_a^*,J_b,J_b^\dag;t]
=\e^{-\mathcal{J}^\dag\mathcal{A}(t)\mathcal{J}}
\mathcal{G}_\mathrm{S}\bm{(}
h^\dag(t)\mathcal{J},\mathcal{J}^\dag h(t)
\bm{)},
\label{eqn:CharFuncSolution}
\end{equation}
where
\begin{equation}
\mathcal{J}^\dag\mathcal{A}(t)\mathcal{J}
=\begin{pmatrix}
J_a^*&J_b^\dag
\end{pmatrix}
\begin{pmatrix}
\mathcal{A}_{aa}(t)&\mathcal{A}_{ab}(t)\\
\mathcal{A}_{ba}(t)&\mathcal{A}_{bb}(t)
\end{pmatrix}
\begin{pmatrix}
J_a\\J_b
\end{pmatrix},\ \ %
h^\dag(t)\mathcal{J}
=\begin{pmatrix}
h_a^*(t)&h_b^\dag(t)
\end{pmatrix}
\begin{pmatrix}
J_a\\J_b
\end{pmatrix}
\end{equation}
with
\begin{subequations}
\label{eqn:ComponentFunctions}
\begin{equation}
\mathcal{A}_{aa}(t)
=\lambda^2\rint_0^t\d t'\rint_0^t\d t''\,
G(t-t')\Phi_{gg}(t',t'')G^*(t-t''),
\end{equation}
\begin{multline}
J_b^\dag\mathcal{A}_{bb}(t)J_b
=\Phi_{J_bJ_b}(t,t)
-2\lambda^2\Re\rint_0^t\d t'\,
(K_{J_bg}*G)(t-t')\Phi_{gJ_b}(t',t)\\
{}+\lambda^4\rint_0^t\d t'\rint_0^t\d t''\,
(K_{J_bg}*G)(t-t')\Phi_{gg}(t',t'')(G^**K_{J_bg}^*)(t-t''),
\label{eqn:Abb}
\end{multline}
\begin{eqnarray}
J_b^\dag\mathcal{A}_{ba}(t)
&=&\lambda\rint_0^t\d t'\,
\Phi_{J_bg}(t,t')G^*(t-t')\nonumber\\
&&{}-\lambda^3\rint_0^t\d t'\rint_0^t\d t''\,
(K_{J_bg}*G)(t-t')\Phi_{gg}(t',t'')G^*(t-t'')
\end{eqnarray}
and
\begin{equation}
h_a(t)
=G(t)+\lambda(G*K_{g(\mathcal{N}\xi)})(t),
\end{equation}
\begin{equation}
J_b^\dag h_b(t)
=K_{J_b(\mathcal{N}\xi)}(t)
-\lambda(K_{J_bg}*G)(t)
-\lambda^2(K_{J_bg}*G*K_{g(\mathcal{N}\xi)})(t).
\label{eqn:hb}
\end{equation}
\end{subequations}
We have introduced
\begin{equation}
K_{fg}(t)=\rint_0^\infty\d\omega\,f_\omega^*\e^{-i\omega t}g_\omega,
\quad
\Phi_{fg}(t,t')
=\rint_0^\infty\d\omega\rint_0^\infty\d\omega'\,
f_\omega^*\e^{-i\omega t}(1+\mathcal{N})_{\omega\omega'}
\e^{i\omega't'}g_{\omega'},
\label{eqn:LocalFunc}
\end{equation}
where $1_{\omega\omega'}=\delta(\omega-\omega')$, and the
convolution
\begin{equation}
(F*G)(t)=\rint_0^t\d t'\,F(t-t')G(t').
\end{equation}
The characteristic functional of the total system
(\ref{eqn:CharFuncSolution}) is exact and valid for any time $t$.

The functions $\lambda(G*K_{g(\mathcal{N}\xi)})(t)$ in $h_a(t)$ and
$K_{J_b(\mathcal{N}\xi)}(t)$ in $h_b(t)$ describe how the initial
correlation propagates, while $\mathcal{A}_{ba}(t)$ and
$\lambda(K_{J_bg}*G)(t)$ in $h_b(t)$ describe the correlation
established through the interaction between system S and reservoir
B\@. System S forgets its initial state through the decay of $G(t)$
and approaches an equilibrium state via the action of
$\mathcal{A}_{aa}(t)$, while $\mathcal{A}_{bb}(t)$ governs the
relaxation of reservoir B into its equilibrium, i.e.~the mixing
state $\Omega_\mathrm{B}$, as explained in the following.

\section{The van Hove Limit of the Characteristic Functional and Discussion}
\label{sec:solution}
We are now in a position to discuss the van Hove limit of the
evolution of the total system and demonstrate the validity of the
general theorem proved in Article I \cite{ref:ArticleI}: (a) the disappearance of the
initial correlation, (b) the factorization of the total system, and
(c) the relaxation into the mixing state, in van Hove's limit.

In order to discuss van Hove's limit, let us remove the (rapid)
oscillation of system S\@. That is, let us look at the
characteristic functional of the density operator $\e^{i
H_\mathrm{S}t}\rho(t)\e^{-i H_\mathrm{S}t}$ in the scaled time
$\tau=\lambda^2t$,
\begin{equation}
\mathcal{G}_\mathrm{I}^{(\lambda)}[J_a,J_a^*,J_b,J_b^\dag;\tau]
=\mathcal{G}[J_a\e^{-i\omega_\mathrm{S}\tau/\lambda^2},
J_a^*\e^{i\omega_\mathrm{S}\tau/\lambda^2},
J_b,J_b^\dag;\tau/\lambda^2].
\label{eqn:CharFuncInt}
\end{equation}
Then, the van Hove limits of the constituent functions
(\ref{eqn:ComponentFunctions}) (Appendix
\ref{app:VanHoveLimitFuncs}),
\begin{subequations}
\label{eqn:ComponentFunctionsLimit}
\begin{eqnarray}
&&\lim_{\lambda\to0}\mathcal{A}_{aa}(\tau/\lambda^2)
=[1+N(\omega_\text{S})](1-\e^{-\Gamma(\omega_\text{S})\tau}),\\
&&\lim_{\lambda\to0}J_b^\dag\mathcal{A}_{bb}(\tau/\lambda^2)J_b
=\rint_0^\infty\d\omega\,
J_{b,\omega}^*[1+N(\omega)]J_{b,\omega},\\
&&\lim_{\lambda\to0}\e^{i\omega_\mathrm{S}\tau/\lambda^2}\mathcal{A}_{ab}(\tau/\lambda^2)J_b
=\lim_{\lambda\to0}J_b^\dag\mathcal{A}_{ba}(\tau/\lambda^2)\e^{-i\omega_\mathrm{S}\tau/\lambda^2}
=0,\\
&&\lim_{\lambda\to0}
h_a(\tau/\lambda^2)\e^{i\omega_\mathrm{S}\tau/\lambda^2}
=\e^{-\Gamma(\omega_\mathrm{S})\tau/2}
\e^{-i\Delta(\omega_\mathrm{S})\tau},\quad
\lim_{\lambda\to0}J_b^\dag h_b(\tau/\lambda^2)=0,
\end{eqnarray}
\end{subequations}
lead us to the van Hove limit of the characteristic functional
(\ref{eqn:CharFuncSolution}),
\begin{eqnarray}
\mathcal{G}_\mathrm{I}[J_a,J_a^*,J_b,J_b^\dag;\tau]
&=&\lim_{\lambda\to0}
\mathcal{G}_\mathrm{I}^{(\lambda)}[J_a,J_a^*,J_b,J_b^\dag;\tau]
\nonumber\\
&=&\e^{-J_a^*J_a[1+N(\omega_\mathrm{S})]
(1-\e^{-\Gamma(\omega_\mathrm{S})\tau})}
\e^{-\rint_0^\infty\d\omega\,J_{b,\omega}^*[1+N(\omega)]J_{b,\omega}}
\nonumber\\
&&{}\times \mathcal{G}_\mathrm{S}(
J_a\e^{-\Gamma(\omega_\mathrm{S})\tau/2}
\e^{i\Delta(\omega_\mathrm{S})\tau},
J_a^*\e^{-\Gamma(\omega_\mathrm{S})\tau/2}
\e^{-i\Delta(\omega_\mathrm{S})\tau}
),\nonumber\\\label{eqn:CharFuncLimit}
\end{eqnarray}
where
\begin{equation}
\Gamma(\omega)=2\pi|g_\omega|^2,\qquad
\Delta(\omega)=\pv\rint_0^\infty\frac{\d\omega'}{2\pi}
\frac{\Gamma(\omega')}{\omega-\omega'}.
\label{eqn:DecayConst}
\end{equation}
It is clear from (\ref{eqn:CharFuncLimit}) that (a) the initial
correlation (or, equivalently, $\xi_\omega$) disappears and (b) the
state of the total system is factorized at all times in van Hove's
limit. Furthermore, (c) the local perturbation in the initial
state $\rho_0$, i.e.~$L$ in (\ref{eqn:EquiSectorModel})
(especially, the contribution of
$\tilde{\mathcal{W}}_{\omega\omega'}$, which appears in the
characteristic functional through
$\tilde{\mathcal{N}}_{\omega\omega'}$), decays out and the
reservoir relaxes into the mixing state $\Omega_\mathrm{B}$ given
in (\ref{eqn:MixingModel}). The dynamics of the system in van
Hove's limit is exactly the same as that derived from the
\textit{uncorrelated} initial state
$\tr\{\rho_0\}\otimes\Omega_\mathrm{B}$ with the mixing state
$\Omega_\mathrm{B}$ and it is actually possible to show that the
density operator $\rho_\text{I}(\tau)$ characterized by the
characteristic functional (\ref{eqn:CharFuncLimit}) obeys the
master equation
\begin{eqnarray}
\frac{\d}{\d\tau}\rho_\mathrm{I}(\tau)
&=&-i[\Delta(\omega_\mathrm{S})a^\dag a,\rho_\mathrm{I}(\tau)]\nonumber\\
&&{}-\frac{1}{2}[1+N(\omega_\mathrm{S})]\Gamma(\omega_\mathrm{S})
[a^\dag a\rho_\mathrm{I}(\tau) +\rho_\mathrm{I}(\tau)a^\dag a
-2a\rho_\mathrm{I}(\tau)a^\dag]\nonumber\\
&&{}-\frac{1}{2}N(\omega_\mathrm{S})\Gamma(\omega_\mathrm{S})
[aa^\dag\rho_\mathrm{I}(\tau) +\rho_\mathrm{I}(\tau)aa^\dag
-2a^\dag\rho_\mathrm{I}(\tau)a].
\end{eqnarray}
This is nothing but the familiar master equation derived from the
factorized initial condition with the reservoir in the thermal
equilibrium state at a finite temperature,
$\rho_0\sim\rho_\mathrm{S}\otimes\e^{-\beta H_\mathrm{B}}$ \cite{ref:WeissTextbook,ref:QuantumNoise}.

These points corroborate the theorem in Article
I, suggesting that the mixing state $\Omega_\mathrm{B}$, which is
contained in the initial state $\rho_0$, should be selected as the
reference state of Nakajima--Zwanzig's projection $\mathcal{P}$.
Note that the characteristic functional in van Hove's limit,
Eq.~(\ref{eqn:CharFuncLimit}), approaches
\begin{equation}
\mathcal{G}_\mathrm{I}[J_a,J_a^*,J_b,J_b^\dag;\tau]
\tolimit{\tau\to\infty}{\to}
\e^{-J_a^*J_a[1+N(\omega_\mathrm{S})]}
\e^{-\rint_0^\infty\d\omega\,J_{b,\omega}^*[1+N(\omega)]J_{b,\omega}},
\end{equation}
which means that the equilibrium state (in van Hove's limit) is
\begin{equation}
\rho_\mathrm{eq}=\frac{1}{Z_\mathrm{eq}}
\e^{-W(\omega_\mathrm{S})a^\dag a}\otimes\Omega_\mathrm{B},\qquad
Z_\mathrm{eq}^{-1}=1-\e^{-W(\omega_\mathrm{S})},
\end{equation}
i.e., system S relaxes into the equilibrium state with the same
structure as that of the mixing state $\Omega_\mathrm{B}$.

As discussed in Article I \cite{ref:ArticleI}, the state of the total system is
factorized through its free evolution and the interaction between
system S and reservoir B is not essential, which is also confirmed
by the present exact solution. In the absence of the interaction,
the exact characteristic functional (\ref{eqn:CharFuncSolution})
reads
\begin{equation}
\mathcal{G}[J_a,J_a^*,J_b,J_b^\dag;t]
=\e^{-\Phi_{J_bJ_b}(t,t)}
\mathcal{G}_\mathrm{S}\bm{(}
J_a\e^{i\omega_\mathrm{S} t}+K_{J_b(\mathcal{N}\xi)}^*(t),
J_a^*\e^{-i\omega_\mathrm{S} t}+K_{J_b(\mathcal{N}\xi)}(t)
\bm{)}
\label{eqn:FreeEvolution}
\end{equation}
and approaches
\begin{equation}
\mathcal{G}[J_a,J_a^*,J_b,J_b^\dag;t]
\tolimit{t\to\infty}{\to}
\e^{-\rint_0^\infty\d\omega\,J_{b,\omega}^*[1+N(\omega)]J_{b,\omega}}
\mathcal{G}_\mathrm{S}(J_a\e^{i\omega_\mathrm{S} t},
J_a^*\e^{-i\omega_\mathrm{S} t})
\label{eqn:CharFuncFreeLimit}
\end{equation}
by Riemann--Lebesgue's lemma [see the discussion below and
Eq.~(\ref{eqn:FreeDecay})]. The state is thus factorized into
\begin{equation}
\rho(t)
\tolimit{t\to\infty}{\to}
\rho_\mathrm{S}(t)\otimes\Omega_\mathrm{B},\qquad
\rho_\mathrm{S}(t)=\e^{\mathcal{L}_\mathrm{S}t}\tr_\mathrm{B}\rho_0
\end{equation}
through the free evolution, which confirms the second part of the
theorem in Article I \cite{ref:ArticleI}.

The timescales of the factorization and the relaxation into the
mixing state are clear from (\ref{eqn:FreeEvolution}): the former is
governed by the function
\begin{equation}
K_{J_b(\mathcal{N}\xi)}(t)
=\rint_0^\infty\d\omega\,\biggl(
\rint_0^\infty\d\omega'\,
J_{b,\omega}^*\mathcal{N}_{\omega\omega'}\xi_{\omega'}
\biggr)\,\e^{-i\omega t}
\label{eqn:FuncFactorization}
\end{equation}
contained in $J_b^\dag h_b(t)$ in (\ref{eqn:hb}), and the latter by
the leading term of $\mathcal{A}_{bb}(t)$ in (\ref{eqn:Abb}),
\begin{eqnarray}
\Phi_{J_bJ_b}(t,t)
&=&\Phi_{J_bJ_b}^{(0)}(0)\nonumber\\
&&{}+2\Re\rint_0^\infty\d\omega\,\biggl(\,\,
\rint_{\omega/2}^\infty\d\bar{\omega}\,
J_{b,\bar{\omega}+\omega/2}^*
\tilde{\mathcal{N}}_{(\bar{\omega}+\omega/2)(\bar{\omega}-\omega/2)}
J_{b,\bar{\omega}-\omega/2}
\biggr)\,\e^{-i\omega t},
\nonumber\\
\label{eqn:ReservoirDecayFourier}
\end{eqnarray}
where
\begin{equation}
\Phi_{fg}^{(0)}(t)
=\rint_0^\infty\d\omega\,f_\omega^*[1+N(\omega)]g_\omega
\e^{-i\omega t}.
\label{eqn:ReservoirDecayNonlocal}
\end{equation}
The timescales of the decay of these functions are determined by the
bandwidths of their Fourier transforms,
\begin{equation}
\tilde{K}_{J_b(\mathcal{N}\xi)}(\omega)
=2\pi\rint_0^\infty\d\omega'\,J_{b,\omega}^*\mathcal{N}_{\omega\omega'}\xi_{\omega'}
\end{equation}
for the former, and
\begin{equation}
\tilde{\Phi}_{J_bJ_b}(\omega)
=2\pi\rint_{\omega/2}^\infty\d\bar{\omega}\,
J_{b,\bar{\omega}+\omega/2}^*\tilde{\mathcal{N}}_{(\bar{\omega}+\omega/2)
(\bar{\omega}-\omega/2)}J_{b,\bar{\omega}-\omega/2}
\end{equation}
for the latter. Therefore, besides the spread of the initial
correlation ($\xi_\omega$) and of the perturbation from the mixing
state ($\tilde{\mathcal{N}}_{\omega\omega'}$), \textit{the size of
the relevant reservoir observables} ($J_{b,\omega}$) influences the
timescales of the factorization and the mixing. In the weak-coupling
regime they are very short compared with the timescale of the
dissipative dynamics of system S, which is of order $1/\lambda^2$ in
the original time $t$.

It is interesting to discuss what happens from a physical point of
view. The initial correlations and the local perturbations propagate
outwards from the region of interest (defined by the ``size'' of the
relevant local observables) and never come back. What remains is the
``unperturbed'' state, that is the mixing state $\Omega_\mathrm{B}$
and is the stable ``ground state'' within the sector
it specifies. The relaxation time of such
a process is the time necessary for the disturbance to pass through
the range of the interaction, that of the initial correlation, and
the extension of the observable. It should be noted that, when we
work in the interaction picture $e^{iH_0t}\rho(t)e^{-iH_0t}$,
instead of $e^{iH_\mathrm{S}t}\rho(t)e^{-iH_\mathrm{S}t}$ considered
above in (\ref{eqn:CharFuncInt}), we should duly take into account the time dependence
of the observables in such a picture, $X(t)=e^{iH_0t}Xe^{-iH_0t}$,
otherwise mixing is not observed.

\section{A Solvable Model with Counter-Rotating Interaction}
\label{sec:model2}
Let us look at another solvable example: the same model as the
previous one (\ref{eqn:SolvableHamiltonian}) but with a different
interaction Hamiltonian
\begin{equation}
H_\mathrm{SB}
=i(a+a^\dag)\rint_0^\infty\d\omega\,
(g_\omega^*b_\omega-g_\omega b_\omega^\dag),
\label{eqn:SolvableHamiltonianC}
\end{equation}
containing \textit{counter-rotating} terms. This model is also
exactly solvable
\cite{ref:FordLewisOConnellQuantumLangevin,ref:Caldeira2000,ref:GrabertReview,ref:CoordinateCouplingModel,ref:OConnell}.
Let us only briefly sketch the main results. More details are
given in Appendix \ref{app:FuncsC}. The exact characteristic
functional of the state of the total system, $\rho(t)$, reads
\begin{eqnarray}
\mathcal{G}[J_a,J_a^*,J_b,J_b^\dag;t]
&=&\e^{-\mathcal{J}^\dag\mathcal{A}(t)\mathcal{J}}
\e^{-\mathcal{J}^\dag\bar{\mathcal{A}}(t)\mathcal{J}^*}
\e^{-\mathcal{J}^\mathrm{T}\bar{\mathcal{A}}^\dag(t)\mathcal{J}}
\nonumber\\
&&{}\times\mathcal{G}_\mathrm{S}\bm{(}
h^\dag(t)\mathcal{J}+\mathcal{J}^\dag\bar{h}(t),
\mathcal{J}^\dag h(t)+\bar{h}^\dag(t)\mathcal{J}
\bm{)}
\label{eqn:CharFuncSolutionC}
\end{eqnarray}
for the same correlated initial state as before, $\rho_0$ in
(\ref{eqn:InitialState}), where $\mathrm{T}$ denotes the transpose
matrix, and
\begin{equation}
\mathcal{J}^\dag\bar{\mathcal{A}}(t)\mathcal{J}^*
=\begin{pmatrix}
J_a^*&J_b^\dag
\end{pmatrix}
\begin{pmatrix}
\bar{\mathcal{A}}_{aa}(t)&\bar{\mathcal{A}}_{ab}(t)\\
\bar{\mathcal{A}}_{ba}(t)&\bar{\mathcal{A}}_{bb}(t)
\end{pmatrix}
\begin{pmatrix}
J_a^*\\J_b^*
\end{pmatrix},\ %
\bar{h}^\dag(t)\mathcal{J}
=\begin{pmatrix}
\bar{h}_a^*(t)&\bar{h}_b^\dag(t)
\end{pmatrix}
\begin{pmatrix}
J_a\\J_b
\end{pmatrix}.
\end{equation}
The details of these functions are given in Appendix
\ref{app:FuncsC}.

This characteristic functional contains different types of terms
from those in the previous example (\ref{eqn:CharFuncSolution}): the
counter-rotating interaction provokes ``squeezing.''\@ In van Hove's
limit, however, these contributions disappear. Indeed, the van Hove
limits of the constituent functions (\ref{eqn:AC})--(\ref{eqn:hC})
of the characteristic functional (\ref{eqn:CharFuncSolutionC}) are
\begin{subequations}
\label{eqn:ComponentFunctionsLimitC}
\begin{eqnarray}
&&\lim_{\lambda\to0}\mathcal{A}_{aa}(\tau/\lambda^2)
=[1+N(\omega_\text{S})](1-\e^{-\Gamma(\omega_\text{S})\tau}),\\
&&\lim_{\lambda\to0}J_b^\dag\mathcal{A}_{bb}(\tau/\lambda^2)J_b
=\rint_0^\infty\d\omega\,
J_{b,\omega}^*[1+N(\omega)]J_{b,\omega},\\
&&\lim_{\lambda\to0}
h_a(\tau/\lambda^2)\e^{i\omega_\mathrm{S}\tau/\lambda^2}
=\e^{-\Gamma(\omega_\mathrm{S})\tau/2}
\e^{-i[\Delta(\omega_\mathrm{S})-\bar{\Delta}(\omega_\mathrm{S})]\tau},
\end{eqnarray}
\end{subequations}
while all other limits vanish (see
Appendix~\ref{app:VanHoveLimitC}), and one ends up with the same
dynamics as the previous one (\ref{eqn:CharFuncLimit}) except for
the frequency shift; $\Delta(\omega_\mathrm{S})$ must be substituted
with $\Delta(\omega_\mathrm{S})-\bar{\Delta}(\omega_\mathrm{S})$,
where $\bar{\Delta}(\omega)$ is defined in (\ref{eqn:ShiftC}). The
present example again supports the validity of the theorem proved in
Article I \cite{ref:ArticleI}: (a) the initial correlation disappears, (b) the state of
the total system is factorized at all times, and (c) the reservoir
remains in the mixing state, in van Hove's limit. The effect of the
counter-rotating interaction manifests itself only in the frequency
shift; no other differences in the resultant dynamics from the
previous example with the rotating-wave interaction
\cite{ref:FacchiPascazioVanHovePhysica}.

Furthermore, the timescales of the factorization and of the mixing
are governed by the functions $K_{J_b(\mathcal{N}\xi)}(t)$
and $\Phi_{J_bJ_b}(t,t)$, respectively 
(see Appendix \ref{app:ComponentsC}); they are the same as those
in the previous example [Eqs.~(\ref{eqn:FuncFactorization}) and
(\ref{eqn:ReservoirDecayFourier})]. This also supports the general
conclusion that the free evolution of the reservoir plays an
essential role for the factorization and the mixing, but the
interaction does not. The counter-rotating interaction gives rise
to no significant effect on the factorization or the mixing.

\section{Concluding Remarks}
\label{sec:Summary}
We have investigated two solvable models in the light of the
general theorem proved in Article I \cite{ref:ArticleI}. In both cases, we confirmed
that when the initial state of the quantum system and the
reservoir is not factorized, a correct application of
Nakajima--Zwanzig's projection method requires that the reference
state of the latter be mixing. In addition, close scrutiny of the
solvable models enabled us to focus on the relevant timescales. It
turns out that an effective factorization of the state of the
total system depends on the \textit{free} dynamics of the
reservoir (responsible for mixing) as well as on the interaction.
Indeed, the free dynamics itself is sufficient to drive a complete
factorization. Moreover, the timescales for mixing (that in turn
govern the very applicability of the projection method in terms of
the ``reference'' state of the reservoir) depend on the ``size''
of local observables of the reservoir: clearly, if one has access
to information that is distributed over \textit{larger} portion of
the reservoir, one can in general detect \textit{finer} deviations
from mixing. The timescales at which Markovianity can be
considered a good approximation depend on the structure of the
local observables that one can measure, that is on the dimension
of the (sub)system whose evolution one wants to describe. This
conclusion, physically sound, is in some sense a strict
consequence of the philosophy at the basis of the $C^*$-algebraic
approach to infinite systems (in the case at hand, the reservoir,
whose observables one can measure).

There are other very interesting problems that we have not
analyzed and that are related to the general features of the
evolutions when it is not permissible to consider a factorized
initial state
\cite{ref:Caldeira2000,ref:GrabertReview,ref:OConnell,ref:Slippage,ref:Oppenheim,ref:Gorini1989,ref:ZoubiAnnPhys2004}.
Among others, the problems related to the (complete) positivity of
the evolution requires additional investigations
\cite{ref:Pechukas,ref:LindbladLendi,ref:Royer,ref:Buzek,ref:Kraus,ref:Benatti,ref:Hanggi2004,ref:SudarshanPRA2004}.
Another interesting issue would be to discuss the applicability of
this method to more articulated (and intriguing) thermodynamical
situations, such as those of nonequilibrium steady states
\cite{ref:Ruelle}, shortly discussed in Article I \cite{ref:ArticleI} (see Fig.~1 in Article I). It
is indeed possible to apply the method we propose to discuss the
relaxation of a system driven by a reservoir at a nonequilibrium
steady state and this aspect will be discussed elsewhere
\cite{ref:DotNESS}.

\ack This work
is partly supported by the bilateral Italian--Japanese Projects
II04C1AF4E on ``Quantum Information, Computation and Communication''
of the Italian Ministry of Instruction, University and Research, and
15C1 on ``Quantum Information and Computation'' of the Italian
Ministry for Foreign Affairs, by the European Community through the Integrated Project EuroSQIP, by the Grant for The 21st Century COE
Program ``Holistic Research and Education Center for Physics of
Self-Organization Systems'' at Waseda University, the Grant-in-Aid
for the COE Research ``Establishment of Molecular Nano-Engineering
by Utilizing Nanostructure Arrays and Its Development into
Micro-Systems'' at Waseda University (No.\ 13CE2003), and the
Grants-in-Aid for Scientific Research on Priority Areas ``Control of
Molecules in Intense Laser Fields'' (No.\ 14077219), ``Dynamics of
Strings and Fields'' (No.\ 13135221), and for Young Scientists (B) (No.\ 18740250) from the Ministry of Education,
Culture, Sports, Science and Technology, Japan, and by Grants-in-Aid
for Scientific Research (C) (Nos.\ 14540280, 17540365, and 18540292) from
the Japan Society for the Promotion of Science.

\appendix
\section{Prototypes of the van Hove Limits}
\label{app:VanHoveLimitFuncs}
The characteristic functional
$\mathcal{G}[J_a,J_a^*,J_b,J_b^\dag;t]$ in
(\ref{eqn:CharFuncSolution}) is expressed in terms of the functions
given in (\ref{eqn:ComponentFunctions}). The van Hove limits of
these functions fall into the following types: by taking the
weak-coupling limit $\lambda\to0$ keeping $\tau=\lambda^2t$ finite,
one obtains
\begin{subequations}
\label{eqn:Prototypes}
\begin{eqnarray}
\text{(i)}&\ &G(t)\e^{i\omega_\text{S}t}
\to\e^{-\Gamma(\omega_\text{S})\tau/2}
\e^{-i\Delta(\omega_\text{S})\tau},\label{eqn:LimitG}\\
\text{(ii)}&&K_{fg}(t)\to0,\qquad \Phi_{fg}(t,t)
\to\rint_0^\infty\d\omega\,f_\omega^*[1+N(\omega)]g_\omega,
\label{eqn:FreeDecay}\\
\text{(iii)}&&(K_{fg}*G*K_{gf'})(t)\e^{i\omega_\text{S}t}
\nonumber\\
&&{}\to\hat{K}_{fg}(-i\omega_\text{S}+0^+)
\hat{K}_{gf'}(-i\omega_\text{S}+0^+)
\e^{-\Gamma(\omega_\text{S})\tau/2}
\e^{-i\Delta(\omega_\text{S})\tau},\label{eqn:VanHoveConv}\\
\text{(iv)}&&\lambda^2\rint_0^t\d t'\rint_0^t\d t''\,
(K_{fg}*G)(t-t')\Phi_{gg}(t',t'')(G^**K_{f'g}^*)(t-t'')\nonumber\\
&&{}\to\hat{K}_{fg}(-i\omega_\text{S}+0^+)
[\hat{K}_{f'g}(-i\omega_\text{S}+0^+)]^*
[1+N(\omega_\text{S})](1-\e^{-\Gamma(\omega_\text{S})\tau}),
\label{eqn:LimitCorrelation}\nonumber\\\\
\text{(v)}&&\rint_0^t\d t'\,
\Phi_{J_bg}(t,t')(G^**K_{fg}^*)(t-t')\nonumber\\
&&{}\to\rint_0^\infty\d\omega\, J_{b,\omega}^*[1+N(\omega)]g_\omega
\frac{[\hat{K}_{fg}(-i\omega+0^+)]^*}{i(\omega-\omega_\text{S})+0^+},\label{eqn:LimitInnerPro}
\end{eqnarray}
\end{subequations}
where  $\Gamma(\omega)$ and $\Delta(\omega)$ are defined in
(\ref{eqn:DecayConst}), and $\hat{K}_{fg}(s)$ is the Laplace
transform of $K_{fg}(t)$ in (\ref{eqn:LocalFunc}). Let us prove
these results.

(i) The van Hove limit of $G(t)$, which is defined in (\ref{eqn:G}),
is the ordinary one
\cite{ref:FacchiPascazioVanHovePhysica}:
\begin{equation}
G(t)\e^{i\omega_\text{S}t}
=\rint_{C_\text{B}}\frac{\d\tilde{s}}{2\pi i}
\frac{\e^{\tilde{s}\tau}}%
{\tilde{s}+\hat{K}(\lambda^2\tilde{s}-i\omega_\text{S})}
\tolimit{\lambda\to0}{\to}
\rint_{C_\text{B}}\frac{\d\tilde{s}}{2\pi i}
\frac{\e^{\tilde{s}\tau}}{\tilde{s}+\hat{K}(-i\omega_\text{S}+0^+)},
\label{eqn:LimitG2}
\end{equation}
which results in (\ref{eqn:LimitG}), by noting the formula for
$\hat{K}(s)$ in (\ref{eqn:G}),
\begin{equation}
\hat{K}(-i\omega_\text{S}+0^+)
=\frac{1}{2}\Gamma(\omega_\text{S}) +i\Delta(\omega_\text{S}),
\end{equation}
with $\Gamma(\omega)$ and $\Delta(\omega)$ defined in (\ref{eqn:DecayConst}).

(ii) The van Hove limits in (\ref{eqn:FreeDecay}) are just the
long-time limits and are due to Riemann--Lebesgue's lemma. The
timescales of the decays are determined by the band widths of their
Fourier transforms. See Eqs.~(\ref{eqn:LocalFunc}) and
(\ref{eqn:ReservoirDecayFourier}).

(iii) In terms of the inverse Laplace transform, the convolution in
(\ref{eqn:VanHoveConv}) is written as
\begin{equation}
(K_{fg}*G*K_{gf'})(t)\e^{i\omega_\text{S}t}
=\rint_{C_\text{B}}\frac{\d s}{2\pi i}
\frac{\hat{K}_{fg}(s)\hat{K}_{gf'}(s)}%
{s+i\omega_\text{S}+\lambda^2\hat{K}(s)}
\e^{(s+i\omega_\text{S})t},
\end{equation}
whose van Hove limit proceeds like in (\ref{eqn:LimitG2}).

(iv) Notice first that the contribution of
$\tilde{\mathcal{N}}_{\omega\omega'}$ to
Eq.~(\ref{eqn:LimitCorrelation}) through the function
$\Phi_{fg}(t,t')$, which represents the effect of the local perturbation $L_\mathrm{B}$ for $\rho_\mathcal{W}$ in
(\ref{eqn:RhoW_OmegaB}), decays out in van Hove's limit, since the
van Hove limit of this contribution is a generalization of (iii) but
with a vanishing prefactor $\lambda^2$ in
(\ref{eqn:LimitCorrelation}).
Therefore, the main contribution comes from the mixing state through
$\Phi_{fg}^{(0)}(t-t')$ defined in
(\ref{eqn:ReservoirDecayNonlocal}): by noting that
\begin{eqnarray}
\lefteqn{\frac{1}{\lambda^2}\Phi_{fg}^{(0)}(t-t')\e^{i\omega_\mathrm{S}(t-t')}}\nonumber\\
&=&\frac{1}{\lambda^2}\rint_0^\infty\d\omega\,
f_\omega^*g_\omega[1+N(\omega)]
\e^{-i(\omega-\omega_\mathrm{S})(t-t')}\nonumber\\
&=&\rint_{-\omega_\mathrm{S}/\lambda^2}^\infty\d\tilde{\omega}\,
f_{\lambda^2\tilde{\omega}+\omega_\mathrm{S}}^*
g_{\lambda^2\tilde{\omega}+\omega_\mathrm{S}}
[1+N(\lambda^2\tilde{\omega}+\omega_\mathrm{S})]
\e^{-i\tilde{\omega}(\tau-\tau')}\nonumber\\
&\tolimit{\lambda\to0}{\to}&\rint_{-\infty}^\infty\d\tilde{\omega}\,
f_{\omega_\mathrm{S}}^*g_{\omega_\mathrm{S}}
[1+N(\omega_\mathrm{S})]\e^{-i\tilde{\omega}(\tau-\tau')}
\nonumber\\
&=&2\pi f_{\omega_\mathrm{S}}^*g_{\omega_\mathrm{S}}
[1+N(\omega_\mathrm{S})]\delta(\tau-\tau'),
\end{eqnarray}
Eq.~(\ref{eqn:LimitCorrelation}) is deduced via
\begin{multline}
\frac{1}{\lambda^2} \rint_0^\tau\d\tau'\rint_0^\tau\d\tau''\,
(K_{fg}*G)\bm{(}(\tau-\tau')/\lambda^2\bm{)}
\e^{i\omega_\text{S}(\tau-\tau')/\lambda^2}
\Phi_{gg}^{(0)}\bm{(}(\tau'-\tau'')/\lambda^2\bm{)}
\e^{i\omega_\text{S}(\tau'-\tau'')/\lambda^2}\\
{}\times(G^**K_{f'g}^*)\bm{(}(\tau-\tau'')/\lambda^2\bm{)}
\e^{-i\omega_\text{S}(\tau-\tau'')/\lambda^2}\\
\tolimit{\lambda\to0}{\to}\hat{K}_{fg}(-i\omega_\text{S}+0^+)
[\hat{K}_{f'g}(-i\omega_\text{S}+0^+)]^*
\Gamma(\omega_\text{S})[1+N(\omega_\text{S})]
\rint_0^\tau\d\tau'\,\e^{-\Gamma(\omega_\text{S})\tau'}.
\end{multline}

(v) While the contribution of $\tilde{\mathcal{N}}_{\omega\omega'}$
decays out in van Hove's limit, which is shown by generalizing (ii)
and (iii), that of $\mathcal{N}_{\omega\omega'}^{(0)}$,
\begin{multline}
\rint_0^\infty\d\omega\,
J_{b,\omega}^*[1+N(\omega)]g_\omega\,\biggl(\,\,
\rint_{C_\text{B}}\frac{\d s}{2\pi i}
\frac{\hat{K}_{fg}(s)}{s+i\omega_\text{S}+\lambda^2\hat{K}(s)}
\frac{e^{(s+i\omega)\tau/\lambda^2}}{s+i\omega}
\biggr)^*\\
=\rint_0^\infty\d\omega\,
J_{b,\omega}^*[1+N(\omega)]g_\omega\,\biggl(\,\,
\rint_{C_\text{B}}\frac{\d\tilde{s}}{2\pi i}
\frac{\hat{K}_{fg}(\lambda^2\tilde{s}-i\omega)}%
{\lambda^2\tilde{s}-i(\omega-\omega_\text{S})+\lambda^2\hat{K}(\lambda^2\tilde{s}-i\omega)}
\frac{e^{\tilde{s}\tau}}{\tilde{s}} \biggr)^*,
\end{multline}
yields (\ref{eqn:LimitInnerPro}).

The prototypes (i)--(v) lead to the van Hove limits of the
components (\ref{eqn:ComponentFunctionsLimit}).

\section{Solution to the Model with the Counter-Rotating Interaction}
\label{app:FuncsC}
We summarize the exact solution to the model with the
counter-rotating interaction (\ref{eqn:CharFuncSolutionC}).

\subsection{Heisenberg Operators}
The exact solution to the Heisenberg equations of motion for
$a(t)=\e^{i Ht}a\e^{-i Ht}$ and $b_\omega(t)=\e^{i
Ht}b_\omega\e^{-i Ht}$ reads
\begin{subequations}
\begin{gather}
a(t)=[F(t)+\lambda^2\bar{F}(t)]a
+\lambda^2\bar{F}(t)a^\dag
+\lambda\rint_0^t\d t'\,F(t-t')[B(t')-B^\dag(t')],\\
b_\omega(t)
=\e^{-i\omega t}b_\omega
-\lambda\rint_0^t\d t'\,\e^{-i\omega(t-t')}g_\omega
[a(t')+a^\dag(t')],
\end{gather}
\end{subequations}
where $B(t)$ is defined in (\ref{eqn:MemoryNoise}) and
\begin{subequations}
\begin{equation}
F(t)=\rint_{C_\mathrm{B}}\frac{\d s}{2\pi i}
\frac{s-i\omega_\mathrm{S}}{s^2+\omega_\mathrm{S}^2+2\lambda^2\omega_\mathrm{S}\hat{L}(s)}\e^{st},
\end{equation}
\begin{equation}
\bar{F}(t)=-\rint_{C_\mathrm{B}}\frac{\d s}{2\pi i}
\frac{i\hat{L}(s)}{s^2+\omega_\mathrm{S}^2+2\lambda^2\omega_\mathrm{S}\hat{L}(s)}\e^{st}
\end{equation}
\end{subequations}
with
\begin{equation}
\hat{L}(s)=-\rint_0^\infty\d\omega\,|g_\omega|^2
\frac{2\omega}{s^2+\omega^2}.
\end{equation}
Note that $F(0^+)=1$, $\dot{F}(0^+)=-i\omega_\mathrm{S}$ and
$\bar{F}(0^+)=0$, $\dot{\bar{F}}(0^+)=0$.

\subsection{Characteristic Functional}
\label{app:ComponentsC}
The characteristic functional of the state of the total system,
$\rho(t)$, is given by (\ref{eqn:CharFuncSolutionC}), which is
composed of the functions
\begin{subequations}
\label{eqn:AC}
\begin{eqnarray}
\mathcal{A}_{aa}(t)
&=&\frac{1}{2}[1-|F(t)+\lambda^2\bar{F}(t)|^2-\lambda^4|\bar{F}(t)|^2]
\nonumber\\
&&{}+\lambda^2\rint_0^t\d t'\rint_0^t\d t''\,
F(t-t')\Re\Phi_{gg}^\beta(t',t'')F^*(t-t''),
\end{eqnarray}
\begin{eqnarray}
J_b^\dag\mathcal{A}_{bb}(t)J_b
&=&\Phi_{J_bJ_b}(t,t)
-\frac{1}{2}\lambda^2[
|(K_{J_bg}*F)(t)|^2
+|(K_{J_bg}^**F)(t)|^2
]\nonumber\\
&&{}+2\lambda^2\Im\rint_0^t\d t'\,
(K_{J_bg}*\Im F)(t-t')\Phi_{gJ_b}^\beta(t',t)\nonumber\\
&&{}+4\lambda^4\rint_0^t\d t'\rint_0^t\d t''\,
(K_{J_bg}*\Im F)(t-t')\nonumber\\
&&\qquad\qquad\qquad\qquad
{}\times\Re\Phi_{gg}^\beta(t',t'')
(\Im F*K_{J_bg}^*)(t-t''),\nonumber\\
\end{eqnarray}
\begin{eqnarray}
J_b^\dag\mathcal{A}_{ba}(t)
&=&\frac{1}{2}\lambda(K_{J_bg}*F)(t)F^*(t)
+i\lambda^3(K_{J_bg}*\Im F)(t)\bar{F}^*(t)\nonumber\\
&&{}+\frac{1}{2}\lambda\rint_0^t\d t'\,
\Phi_{J_bg}^\beta(t,t')F^*(t-t')\nonumber\\
&&{}+2\lambda^3\rint_0^t\d t'\rint_0^t\d t''\,
(K_{J_bg}*\Im F)(t-t')\Im\Phi_{gg}^\beta(t',t'')
F^*(t-t''),\nonumber\\
\end{eqnarray}
\end{subequations}
\begin{subequations}
\label{eqn:AbarC}
\begin{eqnarray}
\bar{\mathcal{A}}_{aa}(t)
&=&\frac{1}{2}\lambda^2[F(t)+\lambda^2\bar{F}(t)]\bar{F}^*(t)\nonumber\\
&&{}-\frac{1}{2}\lambda^2\rint_0^t\d t'\rint_0^t\d t''\,
F^*(t-t')\Re\Phi_{gg}^\beta(t',t'')F^*(t-t''),
\end{eqnarray}
\begin{eqnarray}
J_b^\dag\bar{\mathcal{A}}_{bb}(t)J_b^*
&=&{-\frac{1}{2}}\lambda^2
(K_{J_bg}*F)(t)
(K_{J_bg}*F^*)(t)\nonumber\\
&&{}+i\lambda^2\rint_0^t\d t'\,
\Phi_{J_bg}^\beta(t,t')(\Im F*K_{J_bg})(t-t')\nonumber\\
&&{}+2\lambda^4\rint_0^t\d t'\rint_0^t\d t''\,
(K_{J_bg}*\Im F)(t-t')\nonumber\\
&&\qquad\qquad\qquad\ %
{}\times\Re\Phi_{gg}^\beta(t',t'')
(\Im F*K_{J_bg})(t-t''),
\end{eqnarray}
\begin{eqnarray}
J_b^\dag\bar{\mathcal{A}}_{ba}(t)
&=&-\frac{1}{2}\lambda^3(K_{J_bg}*F)(t)\bar{F}^*(t)
-\frac{1}{2}\lambda\rint_0^t\d t'\,
\Phi_{J_bg}^\beta(t,t')F(t-t')\nonumber\\
&&{}+i\lambda^3\rint_0^t\d t'\rint_0^t\d t''\,
(K_{J_bg}*\Im F)(t-t')\Phi_{gg}^\beta(t',t'')
F(t-t''),\nonumber\\
\end{eqnarray}
\begin{eqnarray}
\bar{\mathcal{A}}_{ab}(t)J_b^*
&=&\frac{1}{2}\lambda[F(t)+\lambda^2\bar{F}(t)]
(K_{J_bg}*F^*)(t)\nonumber\\
&&{}+i\lambda^3\rint_0^t\d t'\rint_0^t\d t''\,
F(t-t')\Phi_{gg}^\beta(t',t'')(K_{J_bg}*\Im F)(t-t''),\nonumber\\
\end{eqnarray}
\end{subequations}
and
\begin{subequations}
\label{eqn:hC}
\begin{equation}
h_a(t)
=F(t)+\lambda(F*K_{g(\mathcal{N}\xi)})(t)+\lambda^2\bar{F}(t),
\end{equation}
\begin{equation}
J_b^\dag h_b(t)
=K_{J_b(\mathcal{N}\xi)}(t)
-\lambda(K_{J_bg}*F)(t)
-2i\lambda^2(K_{J_bg}*\Im F*K_{g(\mathcal{N}\xi)})(t),
\end{equation}
\begin{equation}
\bar{h}_a(t)
=-\lambda(F*K_{g(\mathcal{N}\xi)}^*)(t)-\lambda^2\bar{F}^*(t),
\end{equation}
\begin{equation}
J_b^\dag\bar{h}_b(t)
=-\lambda(K_{J_bg}*F^*)(t)
+2i\lambda^2(K_{J_bg}*\Im F*K_{g(\mathcal{N}\xi)}^*)(t),
\end{equation}
\end{subequations}
where
\begin{equation}
\Phi_{fg}^\beta(t,t')
=\rint_0^\infty\d\omega\rint_0^\infty\d\omega'\,
f_\omega^*\e^{-i\omega t}(1+2\mathcal{N})_{\omega\omega'}
\e^{i\omega't'}g_{\omega'}.
\end{equation}

\subsection{Van Hove's Limit}
\label{app:VanHoveLimitC}
In addition to the prototypes (\ref{eqn:Prototypes}), the following
limits are necessary for the van Hove limit of the characteristic
functional (\ref{eqn:CharFuncSolutionC}): by taking the
weak-coupling limit $\lambda\to0$ keeping $\tau=\lambda^2t$ finite,
we have
\begin{subequations}
\begin{eqnarray}
F(t)\e^{i\omega_\mathrm{S}t}
&=&\rint_{C_\mathrm{B}}\frac{\d\tilde{s}}{2\pi i}
\frac{\lambda^2\tilde{s}-2i\omega_\mathrm{S}}{\lambda^2\tilde{s}^2-2i\omega_\mathrm{S}\tilde{s}+2\omega_\mathrm{S}\hat{L}(\lambda^2\tilde{s}-i\omega_\mathrm{S})}\e^{\tilde{s}\tau}\nonumber\\
&\tolimit{\lambda\to0}{\to}&\rint_{C_\mathrm{B}}\frac{\d\tilde{s}}{2\pi i}
\frac{1}{\tilde{s}+i\hat{L}(-i\omega_\mathrm{S}+0^+)}\e^{\tilde{s}\tau}
=\e^{-\Gamma(\omega_\mathrm{S})\tau/2}\e^{-i[\Delta(\omega_\mathrm{S})-\bar{\Delta}(\omega_\mathrm{S})]\tau},\nonumber\\
\end{eqnarray}
\begin{equation}
F(t)\e^{-i\omega_\mathrm{S}t}
=\rint_{C_\mathrm{B}}\frac{\d\tilde{s}}{2\pi i}
\frac{\lambda^2\tilde{s}}{\lambda^2\tilde{s}^2+2i\omega_\mathrm{S}\tilde{s}+2\omega_\mathrm{S}\hat{L}(\lambda^2\tilde{s}+i\omega_\mathrm{S})}\e^{\tilde{s}\tau}
\tolimit{\lambda\to0}{\to}0,
\end{equation}
\begin{equation}
\lambda^2\bar{F}(t)\e^{\pm i\omega_\mathrm{S}t}
=-\lambda^2\rint_{C_\mathrm{B}}\frac{\d\tilde{s}}{2\pi i}
\frac{i\hat{L}(\lambda^2\tilde{s}\mp i\omega_\mathrm{S})}{\lambda^2\tilde{s}^2\mp2i\omega_\mathrm{S}\tilde{s}+2\omega_\mathrm{S}\hat{L}(\lambda^2\tilde{s}\mp i\omega_\mathrm{S})}\e^{\tilde{s}\tau}
\tolimit{\lambda\to0}{\to}0,
\end{equation}
\end{subequations}
where
\begin{equation}
\pm i\hat{L}(\mp i\omega_\mathrm{S}+0^+)
=\frac{1}{2}\Gamma(\omega_\mathrm{S})
\pm i[\Delta(\omega_\mathrm{S})-\bar{\Delta}(\omega_\mathrm{S})]
\end{equation}
with $\Gamma(\omega)$ and $\Delta(\omega)$ in (\ref{eqn:DecayConst}), and
\begin{equation}
\bar{\Delta}(\omega)
=\rint_0^\infty\frac{\d\omega'}{2\pi}
\frac{\Gamma(\omega')}{\omega+\omega'}.
\label{eqn:ShiftC}
\end{equation}
Then, the van Hove limits of the components
(\ref{eqn:AC})--(\ref{eqn:hC}) of the characteristic functional
(\ref{eqn:CharFuncSolutionC}) yield
(\ref{eqn:ComponentFunctionsLimitC}).




\end{document}